\documentstyle[prd,aps,preprint,floats,epsfig]{revtex}

\tighten

\begin{document}
\draft
 
\pagestyle{empty}

\preprint{
\noindent
\hfill
\begin{minipage}[t]{3in}
\begin{flushright}
LBL--58724 \\
\end{flushright}
\end{minipage}
}

\title{The $X(3872)$ boson: Molecule or charmonium}

\author{
Mahiko Suzuki
}
\address{
Department of Physics and Lawrence Berkeley National Laboratory\\
University of California, Berkeley, California 94720
}


\date{\today}
\maketitle

\begin{abstract}
   It has been argued that the mystery boson $X(3872)$ is a 
molecule state consisting of primarily $D^0\overline{D}^{*0} 
+\overline{D}^0D^{*0}$. In contrast, apparent puzzles and 
potential difficulties have been pointed out for the charmonium 
assignment of $X(3872)$.  We examine several aspects of these 
alternatives by semi-quantitative methods since quantitatively 
accurate results are often hard to reach on them.
We point out that some of the observed properties
of $X(3872)$, in particular, the binding and the production 
rates are incompatible with the molecule interpretation. Despite
puzzles and obstacles,  $X(3872)$ may fit more likely to the 
excited $^3P_1$ charmonium than to the molecule after the 
mixing of $c\overline{c}$ with $D\overline{D}^*+\overline{D}D^*$ 
is taken into account.

\end{abstract}
\pagestyle{plain}
\narrowtext

\setcounter{footnote}{0}
\section{Introduction}

      The narrow state $X(3872)$ was discovered by the 
Belle Collaboration\cite{Belle1} in $B$ decay and subsequently
confirmed by the BaBar Collaboration\cite{BaBar1}. Both the 
CDF\cite{CDF} and the D0 Collaboration\cite{D0} saw a clear
signal of inclusive $X(3872)$ production in $pp$ collision. 
Since $X(3872)$ decays into $\pi^+\pi^-J/\psi$, it was the 
most natural to assign $X(3872)$ to one of the excited 
charmonia\cite{BG,ELQ2}. In fact, the experimental study of the 
$\pi\pi J/\psi$ mode in $B$ decay had been encouraged by theorists, 
notably Eighteen, Lane and Quigg\cite{ELQ}, as a means to explore
the excited charmonia. However, experiment has revealed
unexpected properties of $X(3872)$. 
   
The very narrow width rules out assignment of $X(3872)$ 
to a natural spin-parity state ($P=(-1)^J$ with $C=P$) 
since it would quickly decay into $D\overline{D}$. Among the 
unnatural spin-parity states, the $2^1P_1(h_c')$ charmonium 
of $1^{+-}$ 
was ruled out by Belle\cite{Belle2} at an early stage through 
the angular distribution of the final $\pi^+\pi^-$\cite{PS}. 
If we trust the potential-model calculations of the charmonium 
mass spectrum from the past, there is no suitable candidate 
for $X(3872)$ in the close neighborhood of 3872 MeV. Meanwhile,
the coincidence of the $X(3872)$ mass ($3871.9\pm 0.5\pm 0.5$ 
MeV\cite{Belle1}) with the $D^0\overline{D}^{*0}$ threshold 
($3871.3\pm 0.5$ MeV) prompted many 
theorists\cite{molecule2} to revive the idea of 
molecular states\cite{molecule1} and to speculate that 
$X(3872)$ may be a loosely bound state of $D\overline{D}^*
+\overline{D}D^*$ through a color-neutral force.    

    Then the Belle 
Collaboration\cite{Belle3} reported the startling discovery 
that $X(3872)$ decays into $\omega J/\psi$ as well, actually 
$\pi^+\pi^-\pi^0$ off the $\omega$ resonance peak, with roughly 
the same branching fraction as $\pi^+\pi^-J/\psi$. Since any 
hadron must have a definite charge parity, the dipion $\pi^+\pi^-$ 
in $\pi^+\pi^- J/\psi$ ought to be in $p$-wave ($C=(-1)^l=-1$), 
namely, the $\rho$ and its resonance tail in  $I=1$. In the same 
analysis they saw a signal of the radiative decay mode 
$\gamma J/\psi$\cite{Belle4}, reinforcing the $C=+1$ assignment 
for $X(3872)$. With the mass difference between 
$D^0\overline{D}^{*0}$ and $D^+\overline{D}^{*-}$ being as 
large as 8 MeV, isospin violation may be enhanced if a 
very loosely bound molecule-like state should be formed. Then it
would nicely explain the large isospin mixing in the $X(3872)$ 
decay. The most recent analysis of the decay angular 
distributions\cite{Belle5,Rosner} favors $J^{PC}=1^{++}$ for 
$X(3872)$ among the positive $C$ states. Since $D\overline{D}^*
+\overline{D}D^*$ would be most likely bound in $s$-wave, 
if at all, $1^{++}$ fits well to the molecule.  The molecule 
interpretation has gained steam among some theorists for 
this reason. If $X(3872)$ is indeed a molecule state, it 
would open a large new field in hadron spectroscopy.
Most recently, however, a few pieces of crucial experimental
information\cite{Shen}, though yet preliminary, have appeared 
against the molecule assignment. We are now in a rapidly
moving state of confusion.  

In this paper we cast more doubt on the molecule 
interpretation of $X(3872)$. We first show that unlike the 
proton and the neutron inside a deuteron, no long-range 
potential arises from one-pion exchange between $D$ and 
$\overline{D}^{*}$. Though it may sound strange, 
it is a simple consequence of the numerical accident, 
$m_{D^{*}}-m_{D}-m_{\pi}\simeq 0$, and of the derivative 
$D^*D\pi$ coupling. This makes the ``deuteron-like molecular 
binding'' of $D\overline{D}^* + \overline{D}D^*$ highly unlikely. 
Secondly we argue that the observed decay branching of $B^+\to K^+ 
X(3872)$ in $B$ decay and the inclusive production cross section 
of $X(3872)$ in high-energy proton-proton collision are both 
too large for a loosely bound object of binding energy 1 MeV 
or less.

    We then turn to the charmonium option. We cannot offer a
quantitative resolution as for the mass spectrum computed in 
the potential model. It has been acknowledged that the 
computation involves large uncertainties near and above 
the open charm thresholds. In the 1960's the elaborate 
coupled-channel N/D method was developed in the S-matrix theory 
to compute the spectrum of light hadrons, which were actually 
the molecule states in the present-day language. The N/D method 
in its simple version iterates the Born amplitudes in $s$-channel. 
The computation is very close to the coupled-channel potential 
problem\cite{Collins}. But we were realized at the end how 
difficult it was to obtain reliable quantitative results and 
could not go much beyond semiquantitative analysis in most
cases. The charmonia above the open charm threshold share 
similar uncertainties though the ambiguity coming from the 
high-energy tail is more contained. We leave the difficulty of 
the $X(3872)$ mass with the potential model 
calculation of charmonia to future study. Instead we focus on
the production rate and cross section of $X(3872)$. Our estimate 
shows that because of the very loose binding of the molecule, 
the production rates of the molecule $X(3872)$ should be at 
least an order of magnitude smaller than what we see in experiment. 
In comparison, the charmonium is more easily produced because 
of a little tighter binding. One obvious obstacle to the 
charmonium option is the large isospin breaking in the decay 
$X(3872)\to\rho(\omega)J/\psi$. However, the $\omega$ resonance 
peak is nearly two full widths outside the phase space boundary. 
We ask if there is any chance that this severe kinematic 
suppression of the isospin-allowed decay into $\omega J/\psi$
can bring $B(X\to\rho J/\psi)/B(X\to\omega J/\psi)$ up close 
to $O(1)$. Our finding is that such chance is not ruled out 
within the small range of experimental uncertainty in the 
mass of $X(3872)$. Equally or even more serious is the absolute 
magnitude of the $X(3872)\to\rho(\omega)J/\psi$ decay rate; 
the decay through $c\overline{c}$ annihilation should not 
dominate over the mode $X(3872)\to \rho(\omega)J/\psi$ to
hide them. If the charmonium mixes with a fair amount of the 
$D\overline{D}^*+D^*\overline{D}$ component, it could generate
measurable branching fractions into $\rho(\omega)J/\psi$. 
The best scenario that emerges is as follows: The $X(3872)$ 
state is the $1^{++}$ charmonium that is bound primarily 
with the gluon-exchange force but mixed to $D\overline{D}^*+
D^*\overline{D}$ of $I=0$ through a light quark exchange. 
This scenario will survive even if $X(3872)$ is above the
$D^0\overline{D}^{*0}$ threshold and the decay $X(3872)\to
D^0\overline{D}^{*0}+\overline{D}^0D^{*0}$ indeed 
occurs. We must admit that many of the following analysis  
are only semiquantitative. This is an inevitable shortcoming 
due to the relevant long-distance dynamics that we do not 
really know in details.

\section{Molecule state}
\subsection{One-pion exchange potential}

  We first show that one-pion exchange produces practically 
no force between $D$ and $\overline{D}^*$ contrary to 
intuition. In the limit of $m_{D^*}=m_{D}+m_{\pi}$, there
exists only a $\delta$-function potential in the $s$-wave
channel. Therefore it is incapable of binding $D$ and 
$\overline{D}^*$. 

  Let us define the $D^*D\pi$ coupling by
\begin{equation}
   L_{int} = -ig\overline{D}^{*\mu}\mbox{\boldmath$\tau$}D
       \cdot {\partial}_{\mu} \mbox{\boldmath$\pi$} + h.c.
                \label{coupling} 
\end{equation}  
The value of $g$ can be fixed by the $D^{*+}\to D^+\pi^0$ 
decay rate to $g^2/4\pi=12.8\pm 3.1$. A single pion can be 
exchanged only in the $t$-channel of the $D\overline{D}^*
\to D^*\overline{D}$ scattering. (See Fig. 1.)

\begin{figure}[ht]
\hspace{5cm}
\epsfig{file=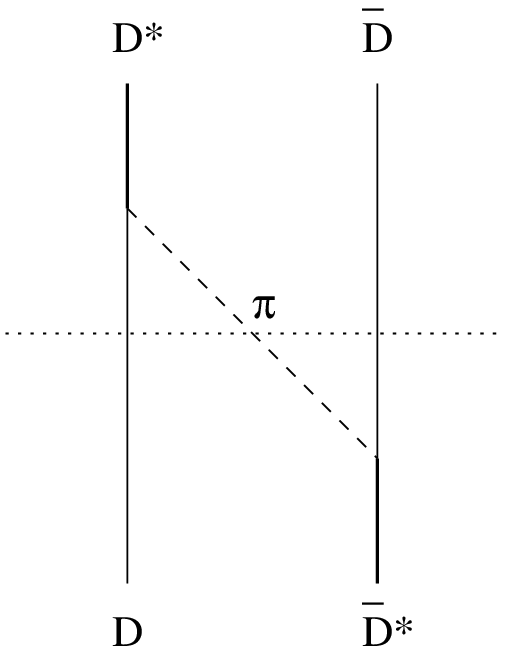,width=4cm,height=5cm}
\caption{The $\pi$-exchange in the scattering
$D\overline{D}^{*}\to D^{*}\overline{D}$.
The intermediate state at the horizontal dotted
line occurs on mass shell for $\pi^0$ exchange.
\label{fig:1}}
\end{figure}

For concreteness, let us examine the $\pi^0$ exchange 
potential 
in $D^0\overline{D}^{*0}\to D^{*0}\overline{D}^0$. The  
one-pion exchange would be normally the primary source of 
the binding force. In the rest of this paper we shall 
write $D\overline{D}^* + \overline{D}D^*$ simply as 
$D\overline{D}^*$ unless it may cause a confusion. 
The potential is extracted from the Born amplitude near 
the threshold: 
\begin{equation}
    T_B = (\mbox{\boldmath$\epsilon$}^{'*}\cdot{\bf q})
              \frac{g^2}{m_{\pi}^2- q^2-i\varepsilon}
          (\mbox{\boldmath$\epsilon$}\cdot{\bf q}),
                     \label{Born} 
\end{equation} 
where ${\bf q}$ is the pion momentum, and
$\mbox{\boldmath$\epsilon$}$ and 
$\mbox{\boldmath$\epsilon$}'$ are the polarizations of 
$\overline{D}^{*0}$ and $D^{*0}$, respectively.
It is important to notice here a peculiarity of kinematics 
in the pion propagator. Since the decay $\overline{D}^{*0}
\to \overline{D}^0\pi^0$ actually occurs with a tiny 
$Q$-value, the denominator of the propagator is 
practically zero as compared with $m_{\pi}$ in the 
nonrelativistic limit;
\begin{eqnarray}
   m_{\pi}^2 - q^2 & \simeq & m_{\pi}^2-\Bigl(m_{D^*}-m_{D}
   +\frac{{\bf q}^2}{8m_{D^*}}-\frac{{\bf q}^2}{8m_{D}}\Bigr)^2
         +{\bf q}^2, \nonumber \\
  &\simeq&  - 2m_{\pi}\Delta +
        \biggl(1+\frac{m_{\pi}^2}{4m_Dm_{D^*}}\biggr){\bf q}^2 
     +O\biggl(\Delta^2,\frac{{\bf q}^2\Delta}{m_D},
     \Bigl(\frac{{\bf q}^2}{m_D}\Bigr)^2\biggr),
                     \label{Deno}                      
\end{eqnarray}
where $\Delta\equiv m_{D^*}-m_{D}-m_{\pi}$ is very small 
($\simeq 7$ MeV). Let us denote the static limit of the
denominator by
\begin{equation}
    \mu^2 = 2m_{\pi}\Delta (\simeq 0.1 m_{\pi}^2).
\end{equation}
With the two powers of ${\bf q}$ from the derivative 
$DD^*\pi$ coupling in the numerator, the Born amplitude 
$T_B$ survives only at ${\bf q}^2\gg 2m_{\pi}\Delta$ and varies
slowly there. The Fourier transform of $T_B$ gives the one-pion 
exchange potential. Since $\overline{D}^{*0}\to\overline{D}^0\pi^0$ 
can occur on mass shell, the principal part of the pion 
denominator is relevant to the potential:  
\begin{eqnarray}
          4m_Dm_{D^*}V({\bf r}) &=& 
              \int {\rm Re}T_B({\bf q})e^{i{\bf q}\cdot{\bf r}}
               \frac{d^3{\bf q}}{(2\pi)^3}\nonumber \\
      &\simeq &\int(\mbox{\boldmath$\epsilon$}^{'*}\cdot{\bf q})
          (\mbox{\boldmath$\epsilon$}\cdot{\bf q})
            \frac{g^2 e^{i{\bf q}\cdot{\bf r}}}{\mu^2-{\bf q}^2}
                    \frac{d^3{\bf q}}{(2\pi)^3}\nonumber \\
      &\simeq& -\frac{1}{3}g^2(\mbox{\boldmath$\epsilon$}^{'*}
               \cdot \mbox{\boldmath$\epsilon$})
               \delta({\bf r}) +  \frac{g^2}{4\pi}
      (\mbox{\boldmath$\epsilon$}^{'*}\cdot\hat{{\bf r}})
      (\mbox{\boldmath$\epsilon$}\cdot\hat{{\bf r}})
      \frac{\mu^2\cos\mu r}{r} \nonumber \\  &+&
      \frac{g^2}{4\pi}\biggl( (\mbox{\boldmath$\epsilon$}^{'*}
               \cdot \mbox{\boldmath$\epsilon$})
         -3(\mbox{\boldmath$\epsilon$}^{'*}\cdot\hat{{\bf r}})
        (\mbox{\boldmath$\epsilon$}\cdot\hat{{\bf r}})\biggr)
     \biggl(\frac{\cos\mu r}{r^3}+\frac{\mu\sin\mu r}{r^2}
         \biggr). \label{pot}
\end{eqnarray}
For the $s$-wave potential relevant to binding of 
$D\overline{D}^*$ into a molecule of $1^{++}$, only a tiny 
contribution of $O(\mu^2)$ survives aside from the 
$\delta$-function term since the last term (valid off $r=0$) 
in Eq. (\ref{pot}) goes away up to $O(\mu^2)$ after 
partial-wave projection.\footnote{
For the $\pi^+$-exchange in $D^0\overline{D}^{*0}\to
D^{*+}D^-$, the Yukawa potential appears rather than the 
oscillatory potential since the on-shell transition does not 
occur with $m_{\overline{D}^{*0}}<m_{D^-}+m_{\pi^+}$. 
However, the leading long-range $s$-wave potential is 
$\sim \mu^2 e^{-\mu r}/r$ instead of $\sim \mu^2\cos\mu r/r$ 
and still totally negligible.}
Contrary to the naive expectation, therefore, the one-pion 
exchange potential is a $\delta$-function in good 
approximation between $D(\overline{D})$ and $\overline{D}^*
(D^*)$. Unlike the one-dimensional $\delta$-function 
potential, the three-dimensional $\delta$-function potential 
cannot generate a bound state when one puts it in the 
Schr\"{o}dinger equation. The Yukawa potential can 
arise only from ``crossed'' multipion exchange. Although the
dipion-exchange potential of an intermediate range plays 
an important role in tightly bound nuclei\cite{Brown} and 
even in the deuteron by providing a spin-isospin independent 
force, lack of
the long range force would not lead to binding of the 
deuteron. Therefore the speculation fails that $X(3872)$ 
may be an analog of the deuteron on the basis of a long range 
force\cite{Torn}. Swanson\cite{Swan} ``appended'' the one-pion
exchange potential with a quark-model potential which he calls 
the ``quark Born diagram potential''\cite{Barnes2}. He turned 
his quark potential into a meson-meson interaction potential 
and added to it the standard one-pion exchange potential of 
range $1/m_{\pi}$, which should be absent between 
$D$ and $\overline{D}^*$ according to our argument. For 
a spatially extended composite state, it is the force of the 
longest range that is most sensitive to binding.  It is not 
clear in his paper how much his quark-model potential is 
responsible for binding of $X(3872)$. While we are unable 
to read quantitative details of calculation in his short paper, 
it appears that some fundamental revision is needed in his 
calculation of binding, at least in the one-pion exchange part. 

The imaginary part of the Born amplitude $T_B$ due to the 
on-shell intermediate state gives an absorptive potential: 
Its $s$-wave contribution is small: ${\rm Im}V(r)=
(g^2\mu^2/12\pi r)(\mbox{\boldmath$\epsilon$}'\cdot
\mbox{\boldmath$\epsilon$})\sin\mu r$. This absorptive
potential is nothing other than a tiny channel coupling 
to $\pi^0 D^0\overline{D}^0$ by pion emission from 
$\overline{D}^{*0}$. The derivative pion coupling or 
the $p$-wave emission of the pion in $D^*\to D\pi$ is the 
source of the behavior ${\rm Im}V(r)\sim \mu^3$ as 
$\mu\to 0$.

It is obvious why the one-pion exchange does not lead to 
the familiar Yukawa potential of $e^{-m_{\pi}r}/r$: 
The pion emission in $D^*\to D+\pi$ can occur without 
violating energy conservation so that the emitted pion can 
travel over long distances. This would otherwise produce 
a long range potential of $\sim 1/r$ and $1/r^3$, but the 
derivative coupling erases them leaving only the 
$\delta$-function in the small $\mu$ limit.

Is it possible to bind $D$ and $\overline{D}^*$ through 
coupling to the $\omega J/\psi$ channel? In general, a
channel coupling can enhance the effective potential for 
the channel of the strongest force. For this channel 
coupling to occur, $D$ (or $D^*$) must be exchanged. (See
Fig. 2.) Since the denominator of the $D(D^*)$ propagator is 
$\simeq 2m_{D/D^*}m_{\omega}$ or larger, the range of this 
off-diagonal Yukawa potential is very short ($\simeq 0.1$ 
fermi). Consequently no strong coupling occurs between the 
$D\overline{D^*}$ and the $\omega J/\psi$ channel at distances 
large enough to be relevant to a loosely bound molecule. 
As for the diagonal potential of $\omega J/\psi$, the elastic 
scattering process is the so-called ``disconnected'' 
quark diagram (Fig. 3) or multi-gluon exchange processes. 
Just as $\phi(s\overline{s})$ interacts only weakly with 
$\pi$, $\rho$, $\omega$ and so forth even at low energies,
elastic $\omega J/\psi$ scattering is expected to be weak and 
hardly a source of binding. To summarize, we see no chance 
of generating a bound state in the $s$-wave $D\overline{D^*}$ 
channels, no matter how the Born diagrams are iterated, 
by solving the Schr\"{o}dinger equation or the Bethe-Salpeter 
equation with or without channel coupling, or by 
approximating the N-function of the matrix N/D with the 
Born amplitudes.

\begin{figure}[ht]
\hspace{5cm}
\epsfig{file=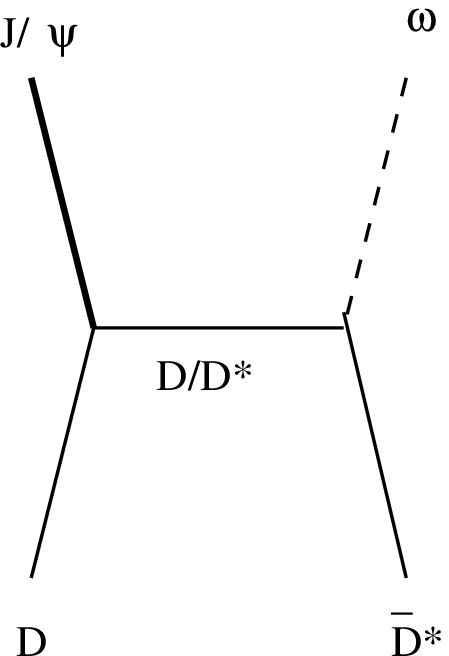,width=3.5cm,height=5cm}
\caption{The $D/D^*$ exchange in $D\overline{D}^*
\to \omega J/\psi$.
\label{fig:2}}
\end{figure}

\begin{figure}[ht]
\hspace{5cm} 
\epsfig{file=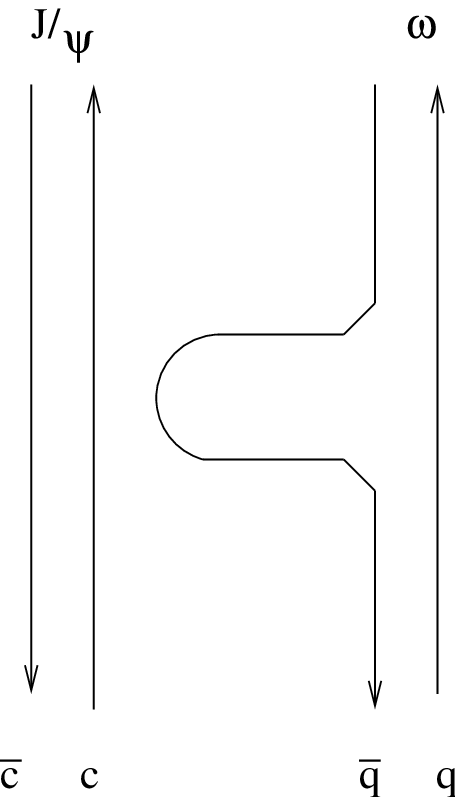,width=3.5cm,height=5cm}
\caption{The disconnected quark diagram of
$\omega J/\psi\to \omega J/\psi$.
\label{fig:3}}
\end{figure}

\begin{figure}[ht]
\hspace{5cm}
\epsfig{file=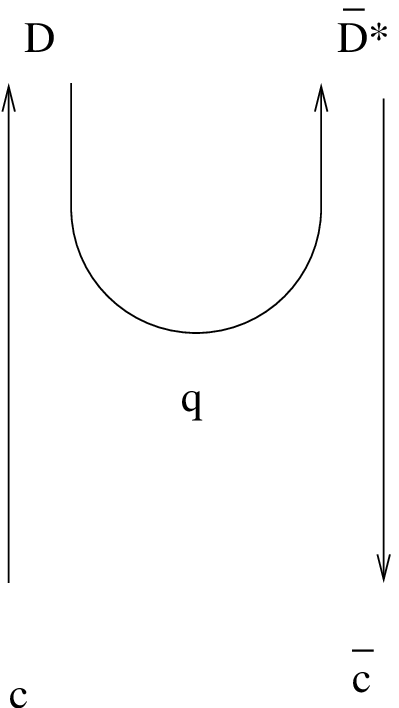,width=3cm,height=5cm}
\caption {The light quark exchange in
$c\overline{c}\to D\overline{D}^*$.
\label{fig:4}}
\end{figure}
  
If there is any relevant channel coupling, it would be to 
the $c\overline{c}$ channel. The confining potential between
$c$ and $\overline{c}$ is stronger than the hadron exchange 
potentials. The coupling between $D\overline{D}^*$ and 
$c\overline{c}$ occurs through the $u(d)$-quark exchange.
(See Fig. 4.) If this channel coupling is strong enough, 
$D\overline{D}^*$ and $c\overline{c}$ can mix substantially. 
Even when such a mixing occurs, the primary source of binding 
is the confining force of gluon exchange between $c$ and 
$\overline{c}$, not the van der Waals force of QCD,
{\em i.e.} not the hadron exchange force in the 
$D\overline{D}^*$ channel. Such a bound state is conceptually 
not the molecule and dynamically different from it . We will 
later discuss this possibility in more detail.

\subsection{Decay rate and production cross section}

 We compare the observed production rates of $X(3872)$ with
the theoretical expectation for the molecule state. We are
limited to semiquantitative discussion here since accurate
calculation would require much more information of
long-distance dynamics and experimental input than we 
have at present. Nonetheless we see a serious difficulty.

 The $X(3872)$ mass has been given by the four experimental 
groups as follows\cite{Belle1,BaBar1,CDF,D0}:
\begin{equation}
      m_{X(3872)} = \left \{ \begin{array}{ll}
               3872.0 \pm 0.6 \pm 0.5 {\rm MeV}& 
                           ({\rm Belle})\nonumber\\
               3873.4 \pm 1.4 {\rm MeV}& 
                           ({\rm BaBar})\nonumber\\
               3871.3 \pm 0.7\pm 0.4 {\rm MeV}& 
                           ({\rm CDFII})\nonumber\\
               3871.8 \pm 3.1 \pm 3.0 {\rm MeV}& 
                           ({\rm D0}).           
       \end{array} \right.
\end{equation} 
Olsen\cite{Belle2} quotes $3871.9\pm 0.5$ MeV
as the weighted average from all groups. We should compare
these numbers with $m_{D^0}+m_{D^{*0}}= 3871.3\pm 1.0$ MeV 
and $m_{D^+}+m_{D^{*+}}= 3879.4\pm 1.0$ MeV\cite{PDG}. 
For $m_D$ and $m_{D^*}$, we have added the quoted errors 
$\pm 0.5$ MeV of the {\em Review of Particle Physics}\cite{PDG} 
since the errors are correlated between $m_{D^0}$ and 
$m_{D^{*0}}$ and between $m_{D^+}$ and $m_{D^{*+}}$. Although 
the central values are on the side of $m_{X(3872)}\geq m_{D^0}
+m_{D^{*0}}$, even the number from Belle that has the highest
accuracy is not conclusive as to whether $m_{X(3872)}$ is 
really above the $D^0\overline{D}^{*0}$ threshold or not.
We have shown the current ranges of $m_{X(3872)}$ and $m_{D^0}+
m_{D^{*0}}$ in Fig. 5. 

\begin{figure}[ht]
\hspace{5cm}
\epsfig{file=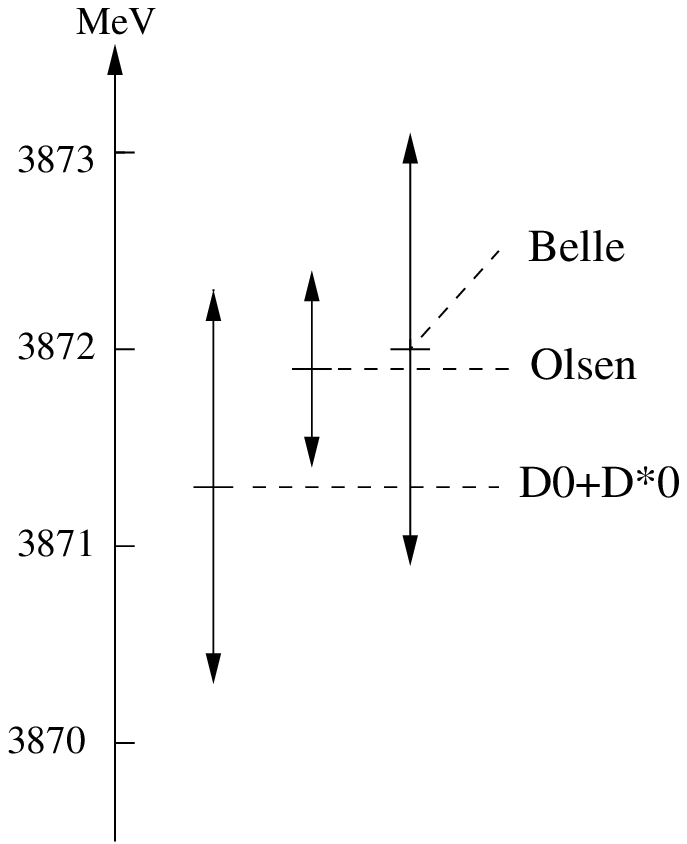,width=5cm,height=5cm}
\caption{The ranges of $m_{X(3872)}$ vs the
value of $m_{D^0}+m_{D^{*0}}$ (denoted as $D0+D^*0$).
\label{fig:5}}
\end{figure}

In order for $X(3872)$ to be a molecular bound state,
$X(3872)$ must be below the $D^0\overline{D}^{*0}$
threshold; $m_{X(3872)} > m_{D0}+m_{D^{*0}}$. However, 
Belle apparently saw a large decay branching to 
$D^0\overline{D}^{0*}$ with the preliminary 
result\cite{Shen},
\begin{equation}
   B(X(3872)\to D^0\overline{D}^0\pi^0)
  /B(X(3872)\to\pi^+\pi^- J/\psi) \approx 10. \label{10}
\end{equation}
If this result holds, the molecule model ought to be 
modified into the ``virtual state'' model\cite{Bugg}. 
For our discussion of the molecule, we proceed with
\begin{equation} 
       m_{X(3872)}< m_{D^0} + m_{D^{*0}}
                  = 3871.3\pm 1.0 {\rm MeV}. \label{mass}
\end{equation}
This places the binding energy of $D^0\overline{D}^{*0}$ 
in $X(3872)$ at less than 1 MeV.

  The production rate of a loosely bound state in 
decay and scattering is obtained generally by 
convoluting the bound state wavefunction with a production 
amplitude of its constituents. When the range of 
production dynamics is much shorter than the size of 
the bound state, the production of the constituents 
and the formation of a bound state factorize. If $X(3872)$ 
is a loosely bound state of $D^0\overline{D}^{*0}$, 
the constituent momenta in the $X(3872)$ rest frame is
$\simeq \sqrt{E_Bm_{D/D^*}}(\simeq 30 {\rm MeV})$. The 
constituents are streaming in parallel with practically 
no relative motion. Even after boosting to the overall 
{\em cm} frame in $B$ decay or in $pp$ collision at 
small rapidity, the relative momentum is still much smaller 
than the characteristic momentum over which the production 
amplitude $A({\bf p}, {\bf k})$ of $D(\frac{1}{2}{\bf p},
{\bf k})\overline{D}^*(\frac{1}{2}{\bf p}-{\bf k},s)$ 
varies significantly.  In this case the invariant amplitude 
of production may be approximated as 
\begin{equation}
   A({\bf p},{\bf k}) \simeq A({\bf p},{\bf 0}),
\end{equation} 
where we have suppressed dependence of $A({\bf p},{\bf k})$ 
on all other variables. Let us express the $s$-wave 
bound state of $\overline{D}D^*$ in the rest frame with
the wavefunction $\tilde{\Psi}({\bf k})$ in momentum 
space and the creation operators $a^{\dagger}_{{\bf k}s}$ 
and $a^{\dagger}_{-{\bf k}}$ of $\overline{D}^*$ and $D$,
respectively:
\begin{equation}
   |X({\bf p}=0,s)\rangle =\int \Psi({\bf k})
            \frac{a^{\dagger}_{{\bf k}s}
          a^{\dagger}_{-{\bf k}}+conj.}{\sqrt{2}}|0\rangle
           \frac{d^3 k}{(2\pi)^3} .
\end{equation}
For production of $X(3872)$ with momentum 
${\bf p}$ and a given helicity, start with the production 
amplitude of $D\overline{D}^*$ with small relative 
momentum ${\bf k}$: 
\begin{equation}
       B/pp \to D + \overline{D}^* + {\rm anything}. 
\end{equation}
Superpose the production amplitude $A({\bf p},{\bf k})$ with 
the bound-state wavefunction first in the {\em cm} frame of 
$\overline{D}D^*$ to get the $X$ production amplitude 
in the molecule rest frame:
\begin{equation}
         \int\frac{1}{\sqrt{4E_DE_{D^*}}}
         \tilde{\Psi}({\bf k})A({\bf 0},{\bf k})
         \frac{d^3 k}{(2 \pi)^3}. \label{convolution}
\end{equation}
When $|{\bf k}|$ is so small that  
$A({\bf 0}, {\bf k})\simeq A({\bf 0},{\bf 0})$, we factor 
out $A({\bf 0},{\bf 0})$ and integrate over ${\bf k}$ 
to obtain $(4m_Dm_{D^*})^{-1/2}
\Psi({\bf 0})A({\bf 0},{\bf 0})$, 
where $\Psi({\bf 0})$ is the wavefunction at 
the origin of the relative position. Moving back to the 
overall {\em cm} frame, one obtains for the production 
rate of $X$ with momentum ${\bf p}$, 
\begin{equation}
  (2\pi)^3E_p\frac{d\Gamma}{d^3p}\simeq
             \frac{|\Psi({\bf 0})|^2}{4m_Dm_{D^*}}
             \sum_i(2\pi)^4\delta^4(P-p+\sum_j p_j)
            |A({\bf p},{\bf 0})|^2, \label{rate} 
\end{equation}
where $P^{\mu}$ is the initial four-momentum and the first 
$\sum$ denotes integration and summation over degrees of 
freedom of all final particles other than $X(3872)$. Because 
of the very loose binding, $|\Psi({\bf 0})|^2$ appears even
when $D$ and $D^*$ are produced by nonlocal interactions.

The wavefunction at origin square $|\Psi({\bf 0})|^2$ in 
Eq. (\ref{rate}) gives us a clue about the production rates. 
The value of $|\Psi({\bf 0})|^2$ is fairly insensitive to 
details of dynamics for a very loosely bound state. We 
estimate as follows:  
Note first model independently that $\tilde{\Psi}({\bf k}) 
\propto 1/(\kappa^2+{\bf k}^2)$ near the $X(3872)$ pole 
($\kappa^2 = 2E_Bm_Dm_{D^*}/(m_D+m_{D^*})$).
The $\tilde{\Psi}({\bf k})$ of this form is not 
suitable for determining $\Psi({\bf 0})$ 
since its Fourier transform
$\Psi({\bf r})\sim e^{-\kappa r}/4\pi r$ is infinite at $r=0$. 
It means that the behavior above $|{\bf k}|\simeq\kappa$ 
matters in determining $\Psi({\bf 0})$. 
$\tilde{\Psi}({\bf k})$ must fall off faster at large 
$|{\bf k}|$ than the simple pole form.\footnote{
If one proceeds to compute Eq. (\ref{convolution}) with
the simple pole form for $\tilde{\Psi}({\bf k})$, there 
is a danger to overestimate the production rate.}
We regularize $\tilde{\Psi}({\bf k})$ into the double
pole form with parameter $M$,
\begin{equation}
     \tilde{\Psi}({\bf k})=
    N\biggl(\frac{1}{{\bf k}^2+\kappa^2}
    -\frac{1}{{\bf k}^2+M^2}\biggr), \label{reg}
\end{equation}
where $N=\sqrt{8\pi\kappa M(M+\kappa)}/(M-\kappa)$ is the 
normalization. Then we obtain
\begin{equation}
 |\Psi({\bf 0})|^2 = \frac{\kappa M(M+\kappa)}{2\pi},
                               \label{worigin}
\end{equation}
For our estimate we choose $M\simeq 3\kappa$ for which the
regulator term contributes only about 5\% at $r=1/\kappa$
in Eq. (\ref{reg}). Then in the ratio to 
$|\Psi({\bf 0})_{\psi(2S)}|^2\simeq 0.024 
{\rm GeV}^3$ of $\psi(2S)$, we obtain
\begin{equation}
 \biggl|\frac{\Psi({\bf 0})_{X(3872)}}{\Psi({\bf 0})_{\psi(2S)}}
     \biggr|^2 \simeq 2.5\times 10^{-3}
 \biggl(\frac{E_B}{0.5 {\rm MeV}}\biggr)^{3/2}. \label{wf} 
\end{equation} 
Since the wavefunction spreads far out in space, the value
at the origin $|\Psi({\bf 0})_{X(3872)}|^2$ is quite small.
Consequently production of $X(3872)$ is highly suppressed in 
$B$ decay and in $pp$ collision. We proceed with Eq. (\ref{wf})
to compare with experiment.

\vskip 0.5cm
\noindent
\underline{$B\to K^+X(3872)$ decay}
 
  For the charmonium production through the decay
$\overline{b}\to\overline{c}c\overline{s}$, 
the color-suppressed tree-interaction is important and 
the factorization calculation appears to be qualitatively 
in line with experiment. The two-body charmonium decay 
$B^+\to K^+(c\overline{c})$ has been observed for $\eta_c$, 
$J/\psi$, $\chi_{c1}$, and $\psi(2S)$\cite{PDG,BaBar2}:
\begin{eqnarray}
        B(B^+\to K^+\eta_c) &=& 9.0\pm 2.7
            \times 10^{-4},\nonumber \\
        B(B^+\to K^+J/\psi) &=& 10.61\pm 0.15\pm 0.48
            \times 10^{-4}, \nonumber \\
        B(B^+\to K^+\chi_{c1}) &=& 5.79\pm 0.26 \pm 0.65
            \times 10^{-4}, \nonumber\\
        B(B^+\to K^+\psi(2S)) &=& 6.49\pm 0.59 \pm 0.97
            \times 10^{-4}.  \label{br}
\end{eqnarray}
In comparison, only the upper bounds have been obtained for 
two-body production of $\chi_{c0}$ and $\chi_{c2}$ that
would not occur in the simple factorization limit:
\begin{eqnarray}
   B(B^+\to K^+\chi_{c0}) &<& 4.4\pm 3.3\pm 0.7
            \times 10^{-4}\nonumber \\
   B(B^+\to K^+\chi_{c2}) &<& 0.09\pm 0.10\pm 0.11
            \times 10^{-4}.
\end{eqnarray}  
The Belle Collaboration\cite{Belle6} has determined the product 
of the branching fractions as
\begin{equation}
     B(B^+\to K^+X(3872))\times B(X(3872)\to\pi^+\pi^-J/\psi)
        = 1.31 \pm 0.24 \pm 0.13 \times 10^{-5}. \label{exp}
\end{equation}
The experimental lower bound can be set on $B(B^+\to 
K^+X(3872))$ with Eq. (\ref{exp}) after taking account of 
the presence of $B(X(3872)\to\pi^+\pi^-\pi^0 J/\psi)$:
\begin{equation}
   B(B^+\to K^+X(3872)) > 2.6 \times 10^{-5}.
          \label{lowerbound}
\end{equation}
However, if the preliminary result of Belle Eq. (\ref{10})
is used, the lower bound rises to
\begin{equation}
    B(B^+\to K^+X(3872)) > 1.6 \times 10^{-4}.
          \label{lowerbound2}
\end{equation}
Computation of the decay amplitude for $B^+\to K^+X(3872)$ in 
the case of the molecular $X(3872)$ has been attempted, but
without a definite numerical result for the branching 
fraction\cite{Braaten1}. It is not surprising when one considers
uncertainties involved in such computations. Nonetheless, if we 
dare to make a very crude estimate, we would proceed with 
Eq. (\ref{rate}) and compare $B(B^+\to K^+X(3872))$ with 
$B(B^+\to K^+\psi(2S))$:
\begin{equation}
 \frac{B(B^+\to K^+X(3872))}{B(B^+\to K^+\psi(2S))}\approx
    \biggl|\frac{A(B^+\to K^+ D\overline{D}^*)}{
          A(B^+\to K^+c\overline{c})}\biggr|^2\times
 \biggl|\frac{\Psi({\bf 0})_{X(3872)}}{
 \Psi({\bf 0})_{\psi(2S)}}\biggr|^2,     \label{r1}
\end{equation}
where the $c\overline{c}$ is in $1^{--}$ and the
$D\overline{D}^*$ is in $1^{++}$, both near their production 
thresholds. We estimate the first factor in the right-hand 
side Eq. (\ref{r1}) with the following reasoning. Imagine 
that a $c\overline{c}$ pair is produced near its threshold, 
most likely in relative $s$-wave. When the $c\overline{c}$ 
invariant mass is below the open charm threshold, the $s$-wave 
$c\overline{c}$ forms a charmonium, $J/\psi$, $\psi(2S)$, 
$\eta_c$, or $\eta'_c$. Above the threshold, they form 
a pair of $D^{(*)}$ and $\overline{D}^{(*)}$ by picking up 
light quarks. Since by assumption the $c\overline{c}$ pair
production amplitude is insensitive to its invariant mass, we
expect 
\begin{equation}
        A(B^+\to K^+(D^{0}\overline{D}^{*0})_{C=+1})\approx   
                   0.5\times  A(B^+\to K^+(c\overline{c})),
\end{equation}
where the energy $E_{c\overline{c}}$ is little below the 
open charm threshold while $E_{D\overline{D}^*}$ is a little 
above it. The factor 0.5 comes from counting of the relevant 
charmonia and the multiplicity of spin, charge, charge parity 
states of charmed meson pairs after $u\overline{u}$ or 
$d\overline{d}$ are picked up.  We use Eq. (\ref{wf}) for 
the second factor in Eq. (\ref{r1}). Combining the two 
factors together, we reach for the molecule $X(3872)$
\begin{eqnarray}
  B(B^+\to K^+X(3872)) &\approx& 1.3\times 10^{-3}
                  \biggl(\frac{E_B}{0.5{\rm MeV}}\biggr)^{3/2} 
                 B(B^+\to K^+\psi(2S)), \nonumber \\
                 &\approx&  0.8\times 10^{-6}
                 \biggl(\frac{E_B}{0.5{\rm MeV}}\biggr)^{3/2}.  
                           \label{molec}
\end{eqnarray}
The number in the last line is more than one order of 
magnitude smaller than the experimental lower bound 
$\simeq 2.6 \times 10^{-5}$ in Eq.(\ref{lowerbound}). 
It is more than two orders of magnitude smaller if
the preliminary result on the $D^0\overline{D}^{*0}$ mode 
is taken into account. Physically speaking, it is not easy 
to produce a large composite object like the molecule 
in the short-distance $B$ decay since $|\Psi({\bf 0})|^2$ 
requires the constituents to come close to each other in
the position space. To enhance the 
production rate to the level of the experimental lower bound, 
$X(3872)$ must be an object of stronger binding. Ironically 
from this viewpoint, if $X(3872)$ were a bound state 
primarily of $D^+D^{*-}$ instead of $D^0\overline{D}^{*0}$, 
the decay branching fraction would be a little 
closer to experiment.

\vskip 0.5cm
\noindent
\underline{$pp\to X(3872)$ + anything}
         
    The CDF Collaboration observed $580\pm 100$ events of $X$ 
production in the region of rapidity $|\eta|<1$ at $\sqrt{s}=
1.96$ TeV when $X(3872)$ is identified with $\pi^+\pi^- J/\psi$. 
They are about 10\% of $\psi(2S)$ production events 
in the same kinematical region\cite{CDF}:
\begin{equation}
  \frac{\sigma(pp\to X(3982) + \cdots)
         B(X(3872)\to\pi^+\pi^- J/\psi)}{
        \sigma(pp\to\psi(2S) + \cdots)
         B(\psi(2S)\to\pi^+\pi^- J/\psi)} =
  \frac{580 \pm 100}{5790\pm 140}\;(\simeq 0.10), \label{br2}     
\end{equation}
where ``$\cdots$'' denotes ``anything''.
The cross section for $\psi(2S)$ above is that of the primary 
production ignoring the $10\sim 20\%$ contribution of the 
feed-down from the $B(\overline{B})$ decay\cite{Bauer}. 
The ratio of the decay branching fractions is 1.6 if $X(3872)
\not{\!\!\!\to}
D^0\overline{D}^{*0}$ and 0.27 if $B(X(3872)\to 
D^0\overline{D}^0\pi^0)= 10\times B(X(3872)\to\pi^+\pi^- 
J/\psi)$ (cf Eq. (\ref{lowerbound2})).
\begin{equation}
   \sigma(pp\to X(3982) + \cdots) 
  = (0.06\sim 0.36)\times\sigma(pp\to\psi(2S) + \cdots).
           \label{cr}
\end{equation} 
We compare the yields for $X(3872)$ and $\psi(2S)$ production 
using the same argument on the amplitude ratio as 
described for the $B$ decay. 
\begin{equation}
       \sigma(p\overline{p}\to X(3872)+\cdots)\approx
 \biggl|\frac{A(p\overline{p}\to D\overline{D}^*+\cdots)}{
   A(p\overline{p}\to c\overline{c}+\cdots)}\biggr|^2
 \biggl|\frac{\Psi({\bf 0})_{X(3872)}}{
    \Psi({\bf 0})_{\psi(2S)}}\biggr|^2
   \times \sigma(p\overline{p}\to\psi(2S)+\cdots).  \label{r2}
\end{equation}
This leads us to the estimate
\begin{equation}
  \sigma(pp\to X(3982) + \cdots)\approx 1.3\times 10^{-3}
       \biggl(\frac{E_B}{0.5 {\rm MeV}}\biggr)^{3/2}
       \times \sigma(pp\to\psi(2S) + \cdots) \label{ratio}
\end{equation}
The ratio of $X(3872)$ and $\psi(2S)$ production is 
one and half to two orders of magnitude smaller than the 
CDF observation, depending on whether $X(3872)$ indeed 
decays into $D^0\overline{D}^{*0}$ or not. 
Strong suppression of the $D\overline{D}^*$ molecule 
production should not be a surprise. Such suppression was 
well known experimentally in similar situations: 
The production cross section of a deuteron ($E_B=2.2$ MeV) 
in $pp\to d\pi^+$ at the {\em cm} energy $\sqrt{s}=2.98$ 
GeV\cite{Smith}, for instance, is only $0.11\pm 0.06$ mb, 
one hundredth of the continuum $pp\to pn\pi^+$ production 
cross section, $11.44\pm 0.65$ mb, at the same energy.

To summarize the case of the molecule, lack of the binding 
force is a serious shortcoming on a theoretical side.  
Despite the numerical uncertainties involved in our 
theoretical estimate, experiment
shows that production of $X(3872)$ is much stronger than 
we expect for a very loosely bound state. Our conclusion 
on the molecule applies to the $D\overline{D}^*$ virtual 
state as well since the wavefunction of the virtual state 
spreads even more than that of the bound state. 

\section{Charmonium}

If $X(3872)$ is a charmonium, the most likely candidate is
the radially excited $^3P_1$ state with $1^{++}$. While 
$^1D_2(2^{-+})$ may look promising, the angular 
distribution analysis disfavors it\cite{Belle5,Rosner}.  
The charmonium interpretation of $X(3872)$ encounters 
two immediate problems.  One is the discrepancy with the
potential-model calculation of the mass 
spectrum\cite{ELQ2,BG,Ka} and the other is the large decay 
branching into the $I=1$ channel, $X(3872)\to\rho J/\psi$. 
We do not attempt to propose any resolution for the potential 
model calculation. Soon after the 
first discovery of charmonia, the mass spectrum below the 
open charm threshold was fitted and reproduced well with the
Cornell potential model\cite{Cornell}. Above the open charm 
threshold, calculation must include coupling to the 
charm-meson-pair channels. While such computation was 
already undertaken even in the early paper of the 
Cornell model, the results involved much larger uncertainties 
than those below the threshold. A decade before
the charmonium, strenuous efforts had been made in 
multi-channel computation of the light hadron mass 
spectrum. However, we were unsuccessful in producing 
quantitatively accurate results and content with the 
semiquantitative predictions in the {\em 
strongest-attractive-force channel} argument.\footnote{
In retrospect, of course, we were barking at a wrong tree
in those days by trying to reproduce the quark-antiquark 
bound states as the ``molecular states''. Nonetheless, 
lack of accuracy or magnitude of uncertainty in the 
multichannel calculation of the bound-state spectrum 
was clearly appreciated. The N/D method was the most
commonly used technique in those computations. It is more 
sophisticated than the naive nonrelativistic potential 
model and its extensions\cite{Collins}.}
With this excuse we will not go into the question as to 
whether or not the channel coupling to $\overline{D}D^*$ 
and $D^*\overline{D}^*$ can indeed lower the $2^3P_1$ mass 
to 3872 MeV from the existing predictions of the potential 
model calculation. As for the production in  $B$-decay and 
$pp$-collision, the larger wavefunction overlap of 
the $1^{++}$ charmonium than that of the molecule enhances
significantly the production rate. However, we are 
unable to produce quantitatively accurate results, since 
the charmonium wavefunction is just as sensitive to the 
channel coupling as the mass spectrum is. Furthermore, the
approximation of the wavefuction-at-origin is less reliable
for the charmonium production in $B$ decay.   

Leaving the mass spectrum aside, we instead focus on the 
issue of the large $\rho J/\psi$ branching fraction.

\subsection {Kinematical suppression of $X(3872)\to\rho J/\psi$}

   Coexistence of the $\omega J/\psi (I=0)$ and $\rho J/\psi
(I=1)$ decay modes\cite{Belle3} appears as a strong argument 
in favor of the molecule model:  
\begin{equation}
 \frac{B(X\to\pi^+\pi^-\pi^0 J/\psi)}{B(X\to\pi^+\pi^- J/\psi)}
     = 1.0 \pm 0.4 \pm 0.3, \label{I}
\end{equation}
However, this relative branching fraction can be misleading 
with respect to the magnitude of isospin mixing in 
$X(3872)$. The $\pi^+\pi^-\pi^0$ in the final state  
$\pi^+\pi^-\pi^0J/\psi$ comes from the far tail of the 
$\omega$ resonance since $\omega J/\psi$ is outside the
phase space boundary. In contrast, $\rho J/\psi$ is right on 
the phase space boundary so that $\pi^+\pi^-$ can come from 
the lower half of the $\rho$ resonance region.  Consequently, 
$\pi^+\pi^-\pi^0$ receives much stronger suppression 
than $\pi^+\pi^-$. Magnitude of this relative suppression 
is purely kinematical and sensitive to the $X(3872)$ 
mass even within the small uncertainty of the $X(3872)$ mass. 

  The mass difference $m_X - m_{J/\psi}$ is smaller than the 
peak value of the $\omega$ resonance ($782.6\pm 0.1$) by 
several MeV:
\begin{equation}
     m_X - m_{J/\psi} - m_{\omega}
        = \left\{ \begin{array}{ll}
                 - 7.5\pm 0.6\pm 0.5 {\rm MeV} &({\rm Belle})\\
                 - 6.1\pm 1.4 {\rm MeV} &({\rm BaBar})\\
                 - 8.2\pm 0.7 {\rm MeV} &({\rm CDF})
                  \end{array} \right..     \label{mass2}
\end{equation}
Recall furthermore that $m_{D^0}+m_{D^{*0}}\geq m_{X(3872)}$ 
is required to prevent the open charm decay, that is, 
\begin{equation}
    m_{X(3872)} \leq m_{D^0}+m_{D^{*0}}= 
         3871.3 \pm 1.0 {\rm MeV}, \label{low}
\end{equation}
which leads to $m_X-m_{J/\psi}-m_{\omega}\leq -8.3\pm 1.0$ 
MeV. The $\pi^+\pi^-\pi^0$ resonates about twice the half 
width $(\frac{1}{2}\Gamma_{\omega}=4.25\pm 0.05$ MeV) 
below the $\omega$ peak. Even at the high-mass end of the
$\pi^+\pi^-\pi^0$ invariant mass, the height of the 
Breit-Wigner resonance is $0.19\sim 0.23$ of its peak value 
for $m_{X(3872)} = 3871.3$ MeV.  Another phase space 
suppression occurs by the relative motion between the 
``off-shell'' $\omega$ and $J/\psi$. The $s$-wave 
threshold factor $|{\bf p}_{J/\psi-\pi\pi\pi}|$ skews the
Breit-Wigner shape and suppresses the high mass end of 
$\pi^+\pi^-\pi^0$. The combined suppression from the two 
sources is quite severe. (See Fig. 6.) According to one 
computation\cite{Braaten2} in the 
molecule model; the observed ratio of Eq. (\ref{I}) would be 
reproduced if the effective $XJ/\psi V$ coupling ratio 
$|g(X\omega J/\psi)/g(X\rho J/\psi)|^2\simeq$ is $11.5\pm 5.5$. 

\begin{figure}[ht]
\hspace{5cm}
\epsfig{file=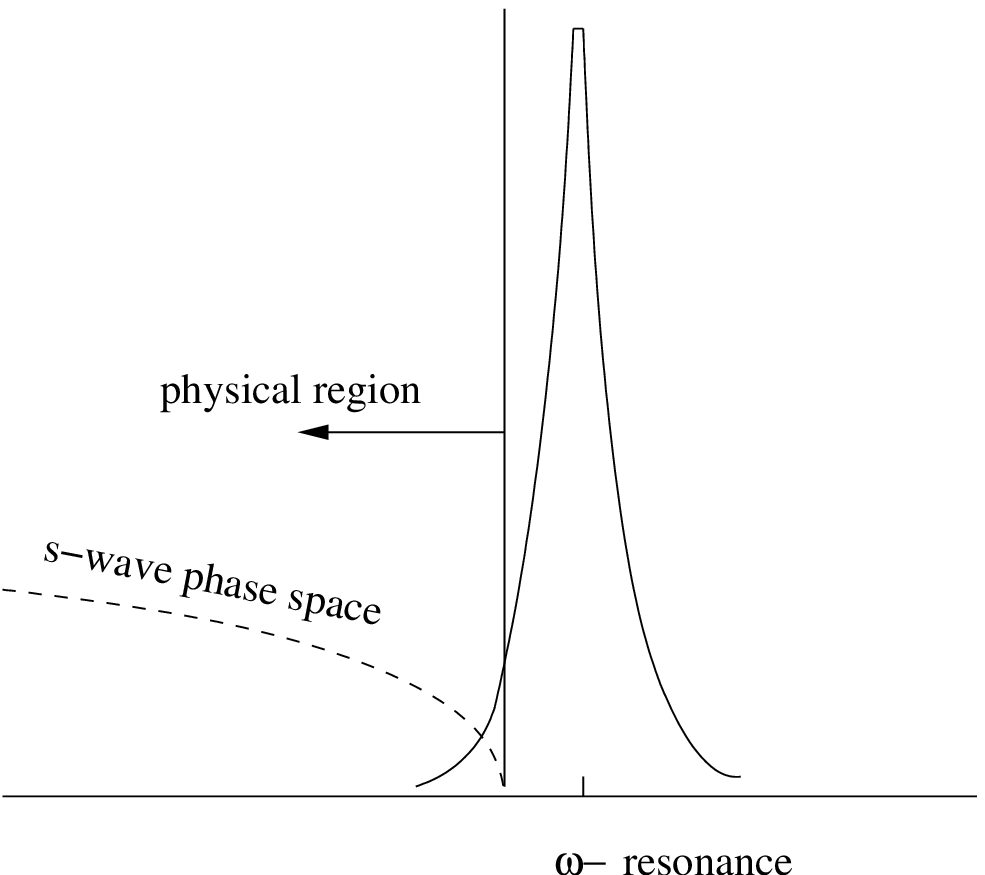,width=6cm,height=5cm}
\caption{ A schematic drawing of
the kinematical suppression near the phase space
boundary for $X(3872)\to \omega J/\psi$. The $\rho$
peak sits right on the phase space boundary with the
height $\simeq 1/18$th of the $\omega$ peak. In this
scale the $\rho$ resonance profile is almost flat.
\label{fig:6}}
\end{figure}

We explore here whether
there is a chance to explain the large isospin breaking of
Eq. (\ref{I}) by the charmonium decay process, not by
the isospin mixing composition of $X(3872)$ that is advocated 
in the molecule model. Assuming that $\omega$ is entirely 
responsible for $\pi^+\pi^-\pi^0$ in line with Belle's 
invariant mass plot, we write out the sensitive part of the
decay rate as   
\begin{equation}
 \Gamma(X(3872)\to\pi^+\pi^-\pi^0J/\psi)
 =\Gamma_0\int_{3m_{\pi}}^{m_{\omega}-\delta m}\frac{|{\bf p}(W)
   |\Gamma_{\omega}}{          
   [(W-m_{\omega})^2+\Gamma_\omega^2/4]}dW, \label{suppression} 
\end{equation}  
where $W$ is the $\pi^+\pi^-\pi^0$ invariant mass, 
${\bf p}(W)$ denotes the three-pion momentum 
${\bf p}_{\pi\pi\pi}$ in the $X(3872)$ rest frame, and
$\delta m \equiv m_{\omega}+m_{J/\psi}-m_{X(3872)}$ is the
distance from the phase space boundary to the $\omega$ peak.
We have computed numerically the integral in 
Eq. (\ref{suppression}) in ratio to the corresponding 
quantity for $\rho$ by varying $m_X$ over 3870 MeV to 
3872 MeV. In carrying this calculation, we have cut off the 
$\pi^+\pi^-\pi^0$ invariant mass at 750 MeV, as chosen 
by Belle, and the $\pi^+\pi^-$ invariant mass at 450 MeV. 
The resulting kinematical suppression $K$ is put in ratio:
\begin{equation}
  \frac{K(\rho J/\psi)}{K(\omega J/\psi)}= 13.3\pm 1.7\;\;\;
               (m_{X(3872)}=3871\pm 1 {\rm MeV}).
\end{equation}
In order to account for the near equality 
of the observed $\rho J/\psi$ and $\omega J/\psi$ rates, 
the production amplitude ratio must be such that  
\begin{equation}
  \biggl|\frac{A(\rho J/\psi)}{A(\omega J/\psi)}\biggr|
     = 0.27\pm 0.02.   \label{sr}
\end{equation}
Being kinematical in origin, this is in general agreement 
with Reference\cite{Braaten2} ($0.27^2$ vs $1/11.3$).
The required number of Eq. (\ref{sr}) is an order of 
magnitude larger than the electromagnetic isospin breaking.  
However it may not be totally out of line for the isospin 
breaking due to the $u$-$d$ quark mass difference, which is 
after all the origin of the 8 MeV mass difference between 
$D^0(D^{*0})$ and $D^+(D^{*+})$. Therefore we next explore 
the magnitude of isospin breaking whether the amplitude 
ratio of Eq. (\ref{sr}) can be realized by the isospin mass 
splitting.

\subsection{Isospin breaking due to mass difference}

  The molecule $X(3872)$ state is predominantly 
made of $D^0\overline{D}^{*0}$ and mixed with 
$D^+\overline{D}^{*-}$, $\omega J/\psi$ and $\rho J/\psi$.
The large $\rho J/\psi$ branching is attributed to the 
isospin mixed composition of $D\overline{D}^{*}$. In the 
charmonium $X(3872)$, the particle composition is almost 
purely in $I=0$ and the isospin breaking occurs during 
the decay process. Although experiment cannot distinguish
between them, the two cases are fundamentally different 
in hadron dynamics. They are not different pictures 
of the same physics related by the quark-hadron duality
or the like.  

   We expect that the main source of the large isospin 
violation is in the $\overline{D}D^*$ intermediate 
states near the thresholds where the splitting 
between $m_{D^0}+m_{D^{*0}}$ and $m_{D^+}+m_{D^{*-}}$ has 
the prominent effect. We can make some quantitative 
discussion of its magnitude with the dispersion relation 
for the $X\to\rho(\omega)J/\psi$ decay amplitudes by 
dispersing the variable $s=p_X^2$. In this way we can at 
least quantify the magnitude within our limited knowledge. 
Since $X(3872)$ is below the $D^0\overline{D}^{*0}$ 
threshold, only the dispersive part exists at 
$s=m_{X(3872)}^2$. Define the invariant decay amplitude 
$A(s)$ by
\begin{equation}
  \langle V(q)J/\psi(p)^{{\rm}out}|
  (\Box+m_X^2)X^{\mu}(0)|0\rangle =\sqrt{\frac{1}{4E_pE_q}}
      \varepsilon^{\mu\nu\kappa\lambda}\epsilon(p)_{\nu}^*
      \epsilon(q)_{\kappa}^*P_{\lambda} A(s) \label{amp}
\end{equation}  
for the decay $X(P)\to \rho/\omega(q)+J/\psi(p)$ with
$P=p+q$ and $s=P^2$. We keep only the $s$-wave term near 
the $D\overline{D}^*$ threshold ignoring the $d$-wave 
contribution for our semiquantitative analysis.
We write the unsubtracted dispersion relation\footnote{
We obviously write an unsubtracted dispersion relation
since a subtracted one would have no predictive power
on the quantity of our interest.} 
for $A(s)$ integrating along the cut which runs on the 
right-hand:
\begin{equation}
 A(s) = \frac{1}{\pi}\int_{s_{{\rm th}}}^{\infty}
      \frac{{\rm Im}A(s')}{s'-s}ds'.
\end{equation}
The lowest intermediate state is $\pi^+\pi^-J/\psi$ 
for $\rho J/\psi$ and $\pi^+\pi^-\pi^0 J/\psi$ for 
$\omega J/\psi$. The dispersion integral of such
intermediate states represents elastic rescattering of 
$\rho(\omega)J/\psi$ in the final state. In the quark 
diagrams they are the ``disconnected processes''
(Fig. 3) that are not the most favored processes. 
The dominant low-energy absorptive parts come from the 
open charm intermediate states $D\overline{D}^* +
\overline{D}D^*$ which couple more strongly with 
$X(3872)$ above the threshold ($s>(m_D+m_{D^*})^2)$. 

Since our primary interest is in the contributions of 
the $D^0\overline{D}^{*0}$ and $D^+\overline{D}^{*-}$
intermediate states of $C=+$ to the dispersion integrals, 
we separate them out as
\begin{eqnarray}
 A_{\omega/\rho}(s) &=& \frac{1}{\pi}
          \int^{\infty}_{(m_{D^0}+m_{D^{*0}})^2}
         \frac{{\rm Im}A(s')_{00}}{s'-s}ds'
  \pm\frac{1}{\pi}\int^{\infty}_{(m_{D^+}+m_{D^{*+}})^2}
             \frac{{\rm Im}A(s')_{-+}}{s'-s}ds' \nonumber \\
 &+& \frac{1}{\pi}\int^{\infty}_{4m_{D*}^2}
         \frac{{\rm Im}A(s')_{**}}{s'-s}ds', 
                 \label{dispersion}
\end{eqnarray}
where the dual sign in front of the second integral is
$+$ for $\omega J/\psi$ and $-$ for $\rho J/\psi$.
Define $s$-wave $XDD^*$ coupling as approximately isospin
invariant as
\begin{equation}
      L_{int} = -m_Xf_{XDD^*}\overline{D}D^{*\mu}X_{\mu},
\end{equation}
and  $D\overline{D}^*\to \rho(\omega) J/\psi$ scattering 
amplitudes as
\begin{eqnarray}
   \langle V(q) J/\psi(p)^{{\rm out}}&|&
            \overline{D}(p_1)D^*(p_2)^{{\rm in}}\rangle \\
      &=& \sqrt{\frac{1}{16E_qE_pE_1E_2}}    
    i(2\pi)^4\delta(q+p-p_1-p_2) \nonumber \\ 
    &\times&\frac{\varepsilon^{\mu\nu\kappa\lambda}}{m_X} 
   \epsilon_{\mu}(q)^*\epsilon_{\nu}(p)^*\epsilon_{\kappa}(p_2)
     [(p+q)_{\lambda}M_1(s,t) + (p-q)_{\lambda}M_2(s,t)].
\end{eqnarray}
The absorptive part for the $\overline{D}D^*$ intermediate 
state is  
\begin{equation}
 {\rm Im}A_{00,-+}(s) = 
   \frac{|{\bf p}(s)_{00,-+}|}{8\pi\sqrt{s}}f_{XDD^*} 
           M(s)_{00,-+}(s)  \label{abs}
\end{equation}
with  $|{\bf p}(s)|=\sqrt{(s-(m_D+m_{D^*})^2)
(s-(m_D-m_{D^*})^2})/2m_X$ and $M(s)_{00,-+}$ is the $s$-wave 
projection of $M_1(s,t)$. The ratio of the $I=1$ to the $I=0$ 
decay amplitude is given by
\begin{equation}
 \frac{A(m_X^2)_{\rho}}{A(m_X^2)_{\omega}}=
   \frac{I_{00}-I_{-+}}{I_{00}+I_{-+}+I_{**}},
\end{equation}
where
\begin{eqnarray}
 I_{00}&=&\frac{1}{\pi}\int_{(m_{D^0}+m_{D^{*0}})^2}
         \frac{{\rm Im}A(s')_{00}}{s'-m_X^2}ds',\nonumber \\
 I_{-+}&=&\frac{1}{\pi}\int_{(m_{D^+}+m_{D^{*+}})^2}
         \frac{{\rm Im}A(s')_{-+}}{s'-m_X^2}ds',\nonumber \\
 I_{**}&=&\frac{1}{\pi}\int_{4m_{D^*}^2}
          \frac{{\rm Im}A(s')_{00}+A(s')_{-+}}{s'-m_X^2}ds'.
              \label{I=1-0}
\end{eqnarray}
 Since the decay $X_{(3872)}\to\rho(\omega)J/\psi$ is a long
distance process, high intermediate states are less important.
That is, the absorptive part ${\rm Im}A(s)$ falls off with
increasing $s$. In order to make a numerical estimate, we need 
to know how far the integrals over $s'$ should be extended. 
Let us choose here the effective cutoff $s_{\rm max}$ no 
higher than $\sqrt{s'}=2m_{D^*} (\simeq m_D+m_{D^*}+m_{\pi})$ 
({\em i.e.,} $\overline{s}_{\rm max}\equiv 
s_{\rm max}-(m_D+m_{D^*})^2\simeq 1 {\rm GeV}^2$ 
and $I_{**}\simeq 0$) and see how large the isospin breaking 
can be. In this energy region, the most important 
$s'$-dependence is in $|{\bf p}(s')|$. We approximate the 
rest of ${\rm Im}A(s')$ to be constant. In this crude 
approximation which is almost independent of dynamics except 
for the value of the cutoff of integral, the ratio of 
Eq. (\ref{I=1-0}) takes a simple form particularly when 
we take the limit of $m_D-m_{D^*}\ll m_D+m_{D^*}$ and
$m_D+m_{D^*}-m_X\ll m_{D^*}-m_D\simeq m_{\pi}$,
\begin{equation}
  \biggl|\frac{A(m_X^2)_{\rho}}{A(m_X^2)_{\omega}}\biggr|
  \approx\frac{\pi\sqrt{m_X/2}[(m_{D^-}+m_{D^{*+}}-m_X)^{1/2}- 
 (m_{D^0}+m_{D^{*0}}-m_X)^{1/2}]}{2\sqrt{\overline{s}_{{\rm max}}}
           -\pi\sqrt{m_X/2}[(m_{D^-}+m_{D^{*+}}-m_X)^{1/2} +
      (m_{D^0}+m_{D^{*0}}-m_X)^{1/2}]}, \label{breaking}
\end{equation}  
For $\sqrt{\overline{s}_{max}}\simeq$1 GeV, the right-hand 
side varies from $\simeq 0.23$ at $m_{X(3872)}=m_{D}+
m_{\overline{D}^{*}}$ to 0.18 at $m_{X(3872)}=m_{D}+
m_{\overline{D}^{*}}-0.8 {\rm MeV}$. These numbers 
are larger than $(m_d-m_u)/m_{\pi}\approx 0.05$. 
The enhancement arises from the fact that the small number 
$m_{D}+m_{D^{*}}-m_X$ is made less small by the square root
threshold factor. Combining this 
isospin breaking with the preceding estimate of 
the $\omega J/\psi$ suppression Eq. (\ref{sr}), we obtain for 
$m_X=3871$ MeV (cf. Eq. (\ref{low}))
\begin{equation}
 \frac{B(X\to\pi^+\pi^-\pi^0 J/\psi)}{B(X\to\pi^+\pi^- J/\psi)}
 \approx \biggl(\frac{1}{0.2}\biggr)^2 
           \times\frac{1}{13.3} \simeq 2\;\; 
           (vs\; 1.0\pm 0.4 \pm 0.3). \label{breaking2}
\end{equation}
We are short of the central value of experiment by 
factor two with large uncertainties. The ratio
$|A(m_X^2)_{\rho}/A(m_X^2)_{\omega}|$ is sensitive 
to the value of the cutoff $s_{{\rm max}}$. If 
$\sqrt{\overline{s}_{\rm max}}=\sqrt{s_{\rm max}-
(m_D+m_{D^*})^2}$ is lowered from 1 GeV to 700 MeV, 
the ratio of Eq. (\ref{breaking2}) would become unity. 
Although choosing the cutoff $\sqrt{s_{{\rm max}}}$ 
below $2m_{D^*}$ would be unrealistic, tapering off 
${\rm Im}A_{00,-+}(s')$ towards $\sqrt{s_{{\rm max}}}
=2m_{D^*}$ has the same effect.

   We have computed above only the kinematical effect 
due to the mass splitting between the charged and 
neutral charmed mesons. We must admit that we have 
stretched the numbers to the limit in this estimate.
Since we have used the isospin symmetric $XDD^*$ coupling, 
however, our $(I_{00}-I_{-+})$ term does not include 
the final state rescattering of $D\overline{D}^*$ 
in $I=1$ channel at $m_{X(3872)}^2<s'<s_{\rm max}$. It
can generate additional enhancement or suppression to
the integrals. But we do not have enough knowledge to 
analyze such dynamical effects.

We should learn from the exercise above that although 
the problem of the large isospin breaking is serious, 
it is too early to reject the charmonium interpretation 
of $X(3872)$ on the basis of the large $\rho J/\psi$ 
decay branching alone.

\subsection{Magnitude of $\rho(\omega)J/\psi$ rates}

   We should be equally or even more concerned with the 
magnitude of the branching fractions themselves than 
their ratio. In order to observe the decay mode 
$\rho(\omega) J/\psi$ in experiment at all, it must have 
a branching fraction large enough to stand out of the 
annihilation process $^3P_1\to ggg (q\overline{q}g)\to 
hadrons$. With the kinematical suppression being so strong, 
would the decays into $\rho(\omega)J/\psi$ be still visible ?  
This problem made some theorists suspicious about the 
charmonium interpretation of $X(3872)$ already at an early 
stage\footnote{
See for instance Reference \cite{BG}.} We have no means to 
estimate reliably the magnitude of the coupling $f_{XDD^*}$ 
nor the amplitude ${\rm Im}A(s)_{00,-+}$ in Eq. (\ref{abs}).
Nonetheless we must address to this question.

  Since we mean by charmonium the $c\overline{c}$ state 
that is bound primarily by the confining force, mixing of
$c\overline{c}$ with $D\overline{D}^*$ is not large in a
``clean'' charmonium by definition. Therefore 
$|\Psi'(0)_{X(3872)}|^2$ should not be very far from 
the value of the simplest potential model that ignores 
the open charm channels.  When we take the ratio of the 
branching fractions in $B$ decay to that of the production 
cross sections in $pp$ collision for $X(3872)$ and 
$\psi(2S)$, the wavefunctions $|\Psi(0)_{\psi(2S)}|^2$ and 
$|\Psi'(0)_{X(3872)}|^2$ cancel out. It is a reasonable
assumption that the ratio of the $c\overline{c}$ production 
in $s$-wave to $p$-wave in $B$ decay is similar to the 
same ratio in $pp$ collision at low rapidity. If so, 
we obtain 
\begin{equation}
   \frac{B(B\to K^+X(3872))}{B(B\to K^+\psi(2S))}
 \approx \frac{\sigma(pp\to X(3872)+{\rm anything})}{
           \sigma(pp\to \psi(2S)+{\rm anything})}.        
\end{equation} 
Denoting the branching fraction $B(X(3872)
\to\pi^+\pi^-J/\psi)$ by $b_{\pi^+\pi^-}$, we                                   
have $0.02/b_{\pi^+\pi^-}$ for the left-hand side with Eqs. 
(\ref{br}) and (\ref{exp}) while $0.03/b_{\pi^+\pi^-}$ for 
the right-hand side with Eq. (\ref{br2}). The both sides 
are in fair agreement with each other, which may indicate 
that our crude reasoning on the long-distance physics is 
not out of line. However, we should be concerned with
the annihilation decay $X(3872)\to ggg + q\overline{q}g$. 
This mode, which is insensitive to the $D$ and $D^*$ masses, 
is expected to be two and half orders of magnitude stronger 
than $2^3P_1\to\gamma J/\psi$ when the mass is adjusted to 
3872 MeV in the estimate by Barnes and Godfrey\cite{BG}. 
Since $B(X(3872)\to\gamma J/\psi)$ is about 10\% of 
$B(X(3872)\to\omega J/\psi)$ according to Belle, 
$b_{\pi^+\pi^-}$ should be quite small:
\begin{equation}
      b_{\pi^+\pi^-} \simeq 0.015  \label{A} 
\end{equation}
with the annihilation mode
even in the absence of the $D^0\overline{D}^{*0}$ mode.
For the pure charmonium $X(3872)$, Eq. (\ref{A}) 
requires with Eq. (\ref{exp}) that $\psi(2S)$ and $X(3872)$ 
are equally copiously produced in $B$-decay in the case of 
$X(3872)\not{\!\!\to}D^0\overline{D}^{*0}$. 
$X(3872)$ could be more strongly produced than $\psi(2S)$ 
if $X(3872)$ indeed decays into $D^0\overline{D}^{*0}$. 
This is a serious problem: $|\Psi'(0)|^2$ must be large 
enough for sufficient production of $X(3872)$ in $B$ decay 
and $pp$ collision while a large value of $|\Psi'(0)|^2$ 
potentially makes the annihilation decay too strong and 
the decay $X(3872)\to\pi^+\pi^-(\pi^0)J/\psi$ invisible. 
One escape from this problem is that $D\overline{D}^*$ 
is already present in $X(3872)$ and easily decays into 
$X(3872)\to\pi^+\pi^-(\pi^0)J/\psi$ by the quark 
rearrangement process. We now look into the $D\overline{D}^*$
content of the charmonium, namely mixing with
$D\overline{D}^*$.

\section{Large mixing between charmonium and charmed mesons}    

   The channel coupling between $c\overline{c}$ and 
$D^{(*)}\overline{D}^{(*)}$ was studied in the potential 
model. The s-wave channels of $c\overline{c}$ were studied
numerically even in the expanded paper of the
Cornell model\cite{Cornell}. However, it is not clear how 
much numerical uncertainty should be attached to the mass 
spectrum involving $c\overline{c}$ at and above the open 
charm threshold. Diagonalization must be made not simply 
for the multichannel amplitudes at the energy of a bound 
state, but for certain integrals of them.  In the 
multi-channel N/D method, the tractable approximation
close to the potential model is to represent the N-function
of given $J^{PC}$ by the Born amplitudes and to set to 
$D(\infty)=1$ for the D-function. One advantage of the
N/D method over the potential model is its notational 
simplicity in discussing channel coupling. Normalize the 
partial-wave amplitude such that unitarity holds as 
${\rm Im}a_J(s)=a_J^{\dagger}(s)\rho(s)^{-1}a_J(s)$ and
${\rm Im}[a_J(s)^{-1}]=-\rho(s)$ where $\rho(s)$ is the 
diagonal phase space matrix\cite{Collins,BJ}. Introduce 
the N and D functions by $a_J(s)=N_J(s)D_J(s)^{-1}$. 
Suppressing the subscript $J$ hereafter,  
\begin{equation}
     N(s)=\left( \begin{array}{ll}
              B_{11}(s) & B_{12}(s) \\
              B_{21}(s) & B_{22}(s) \end{array} \right),
                \label{N}
\end{equation}
where the rows and columns refer to $c\overline{c}$ and
$D\overline{D}^*$ of $I=0$. We do not include other
channels such as $\omega J/\psi$ and $ggg$ as 
constituents of $X(3872)$ here since coupling of
$c\overline{c}$ to these channels is expected to be
much weaker than to $D\overline{D}^*$. We treat the
coupling to the weakly coupled channels as decay. 
Such separation would become less clear if their channel 
coupling were stronger. The $(2\times 2)$ D-function is 
\begin{equation}
    D(s) = I -\frac{1}{\pi}\int_{s_{{\rm th}}}^{\infty}
           \frac{\rho(s')N(s')ds'}{s'-s}. \label{D}
\end{equation}
The zero of ${\rm det}D(s)$ gives the mass square of 
a bound state and the diagonalization matrix of $D(s)$
at the zero determines the composition of the bound 
state.\footnote{
In the case of a single channel Eqs. (\ref{N}) and 
(\ref{D}) combined reduce to solving in effect the 
corresponding Bethe-Salpeter equation where the kernel 
is the Born amplitude. In the nonrelativistic limit, 
therefore, it is equivalent to the potential calculation.}
Although it is hard to get numerical results in our case, 
we can make one simple observation about the 
$c\overline{c}$-$D\overline{D}^*$ mixing. 

We are interested in the possibility that a large 
off-diagonal element $B_{12}(s) (=B_{21}(s))$ for 
$c\overline{c}\leftrightarrow D\overline{D}^*$ causes 
a strong mixing. The Born diagram for $c\overline{c} 
\leftrightarrow D\overline{D}^*$ is a light quark 
exchange (Fig. 4). We have already shown in Section II 
that there is practically no force between $D(D^*)$ 
and $\overline{D^*}(\overline{D})$ in the $1^{++}$ channel.  
Therefore $B_{22}(s)\simeq 0$ in the low-energy region 
and $D(s)$ has the pattern of 
\begin{equation}
  D(s) \simeq \left( \begin{array}{cc}
     1+\overline{D}_{11}(s) & \overline{D}_{12}(s) \\
     \overline{D}_{21}(s) &    1    \end{array} \right),
\end{equation}
where $\overline{D}_{ij}(s)=-\int[\rho N/(s'-s)]_{ij}ds'$.
It is diagonalized by the orthogonal rotation of angle
$\theta$ that is given by
\begin{equation}
   \tan 2\theta = \frac{2\overline{D}'_{12}(m_X^2)}{
       \overline{D}'_{11}(m_X^2)},
\end{equation}     
where the prime on $\overline{D}_{ij}$ denotes the
first derivative in $s$. 
The confining potential of gluon exchange for 
$c\overline{c}\leftrightarrow c\overline{c}$ is no
weaker than the quark exchange for 
$c\overline{c}\leftrightarrow  D\overline{D}^*$. We
can thus set a bound on the mixing,
\begin{equation}
  \tan 2\theta < 2 \;\;\; \to \;\;\; |\theta|< 32^{\circ}.
\end{equation}   
If the mixing of $\theta\simeq 32^{\circ}$ really occurs, 
the binding force would be enhanced by about 60\% as 
a channel coupling effect.

 We make the following observation from this exercise:
It is possible for the $1^{++}$ charmonium to contain
a $D\overline{D}^*$ component up to one third 
($\simeq(\tan32^{\circ})^2$). A stronger mixing is 
possible only if the force of shorter distances in the 
elastic $D\overline{D}^*$ channel should play a role in binding.
The mass splitting between $D^{(*)0}$ and ${D^{(*)+}}$ 
should not generate a very large departure from isospin 
symmetry in the $D\overline{D}^*$ content of $X(3872)$ 
since the effective binding force is the integral of 
the Born amplitudes smoothed out over energy, not 
the Born amplitudes themselves at or near $m_{X(3872)}$. 
The $D\overline{D}^*$ content of $X(3872)$ is  
approximately in $I=0$. It should also be pointed out that
the binding force due to the channel coupling, determined 
by $D_{12}(m_X^2)$ is insensitive whether $X(3872)$ is 
above or below the $D^0\overline{D}^{*0}$ threshold.

 In the presence of a large mixing, production of $X(3872)$
occurs mainly through the dominant $c\overline{c}$ 
component. The production is robust since $|\Psi'(0)|^2$ 
is large for the p-wave charmonia. On the other hand the 
$D\overline{D}^*$ component is unimportant for production
since its $|\Psi({\bf 0})|^2$ is small. However, the
$D\overline{D^*}$ component plays the major role in the 
decay into $\pi^+\pi^-J/\psi$ and $\pi^+\pi^-\pi^0J/\psi$ 
since the virtual $D\overline{D}^*$ component can decay
more easily into those channels than $c\overline{c}$ 
does. The decay $D\overline{D}^*\to \rho(\omega) J/\psi$ 
is a quark-rearrangement process and the strength of the
$D\overline{D}^*$ binding is unimportant. Since the 
$D\overline{D}^*$ component can make up to one third, 
$X(3872)\to\rho(\omega)J/\psi$ is more competitive with 
$X(3872)\to ggg + q\overline{q}g$ than in the case of 
the unmixed pure charmonium. In our dispersion relation of 
the tree-point function in Section III, a large 
$D\overline{D^*}$ component is present when the coupling 
$f_{XDD{^*}}$ is strong and the transition to the
$D\overline{D}^*$ intermediate state is easy. The $2^3P_1$
charmonium mixed with $D\overline{D}^*$ was discussed by
Meng, Gao and Chao\cite{MGC} in the case that binding is 
entirely due to the confining force.

\section{Concluding remarks}

   We have examined the molecule model and the charmonium 
model for $X(3872)$.  The main motivations of the
molecule idea are the coincidence of the $X(3872)$ mass
with $m_{D^0}+m_{D^{*0}}$ and the large isospin violation
in the decay modes. However, there is no long range force 
to bind $D$ and $\overline{D}^*$ into a deuteron-like state. 
The observed production rates of $X(3872)$ in $B$ decay 
and $pp$ collision are too large for a very loosely bound 
state. On the other hand the charmonium has its share 
of difficulties; The mass does not agree with the potential 
model prediction of the $2^3P_1$ state and the large decay 
branching for $c\overline{c}\to ggg+q\overline{q}g$ could 
make the experimental signal of $\pi^+\pi^-(\pi^0)J/\psi$
hardly visible.  One resolution appears to be as
follows: $X(3872)$ is bound primarily by the confining 
force between $c$ and $\overline{c}$ which is boosted by
the channel coupling to $D\overline{D}^*$. Production of
$X(3872)$ occurs mostly through its $c\overline{c}$ 
component. The $D\overline{D}^*$ component is in $I=0$ 
in good approximation. The large isospin breaking of 
the decay mode $\rho J/\psi$ relative to $\omega J/\psi$ 
results from a normal magnitude of isospin breaking due
to $D^{(*)+}-D^{(0)*}$ mass difference that is
enhanced by the severe kinematical suppression of the 
$\omega J/\psi$ mode. This picture is very different
conceptually from the molecule model or the multiquark 
model. In this picture, the binding force comes 
primarily from $c\overline{c}$ and secondarily from
the channel coupling. The elastic $D\overline{D}^*$ 
channel provides practically no binding force. This is 
an important distinction from the viewpoint of hadron
spectroscopy because if our picture is right, $X(3872)$ 
will not an opening of a flood gate for multiquark or 
molecule states.

   How can we distinguish among the different models and
pictures by experiment ?  We should test the particle 
content of $X(3872)$. In the case of the charmonium and 
the charmonium mixed with $D\overline{D}^*$ of $I=0$, 
the production rate is the same for $B^+$ and $B^0$ by 
isospin symmetry as long as it occurs through the
dominant interaction $\overline{b}\to \overline{c}c
\overline{s}$. Independent of dynamics, therefore, we 
expect for the charmonium $X(3872)$,
\begin{equation}
 \Gamma(B^+\to K^+X(3872)) = \Gamma(B^0\to K^0+X(3872)).
\end{equation}
This equality should hold equally well in the case of the 
large mixing between $c\overline{c}$ and $D\overline{D}^*$ 
since $X(3872)$ is produced primarily through its 
$c\overline{c}$ component whose $|\Psi'(0)|^2$ is large. 
For the molecule $X(3872)$ where the $D\overline{D}^*$ 
component has a sizable isospin breaking, the equality 
would be violated since the decay amplitudes into $I=0$ 
and 1, or $D^0\overline{D}^{0-}$ and $D^+\overline{D}^{*-}$, 
are dynamical independent. If the molecule $X(3872)$ 
is made as
\begin{equation}
    X(3872) = D^0\overline{D}^{*0}\cos\alpha +
               D^+\overline{D}^{*-}\sin\alpha,
\end{equation} 
we can express the decay amplitudes for $B^+\to K^+X(3872)$ 
as
\begin{equation}
    A(B^+\to K^+X(3872)) = A_{00}\cos\alpha +
                            A_{+-}\sin\alpha. 
\end{equation}
Then the decay amplitude for $B^0\to K^0X(3872)$ is obtained 
by isospin rotation:
\begin{equation}
    A(B^0\to K^0X (3872)) = - A_{+-}\cos\alpha -
                            A_{00}\sin\alpha.
\end{equation}
The color of the spectator quark ($u/d$) matches that of 
the $\overline{c}$-quark from $\overline{b}\to\overline{c}c
\overline{s}$ in $B^+\to K^+ D^0\overline{D}^{*0}$ and 
$B^0\to K^0 D^+\overline{D}^{*-}$, but not in
$B^+\to K^+ D^+\overline{D}^{*-}$ nor
$B^0\to K^0 D^0\overline{D}^{*0}$.  
In the $1/N_c$ expansion, therefore, 
$A_{+-}=O(1/N_c)\times A_{00}$. Consequently for 
the molecule $X(3872)$ we have
\begin{equation}
    A(B^0\to K^0X(3872)) = 
            -\biggl(\tan\alpha +O(1/N_c)\biggr)
     \times A(B^+\to K^+X(3872)). 
\end{equation}
Since $D^0\overline{D}^{*0}$ dominates over
$D^+\overline{D}^{*-}$ in the molecule 
($\tan^2\alpha < 1$), we predict for the molecule
\begin{eqnarray}
     \Gamma(B^0\to K^0X(3872)) &\simeq& \tan^2\alpha\; 
            \Gamma(B^+\to K^+X(3872)) \nonumber \\
             &<& \Gamma(B^+\to K^+X(3872)).
\end{eqnarray}
Comparison of the $B^0$ decay with the $B^+$ decay clearly 
distinguishes between the deuteron-like 
bound state of $D^0\overline{D}^{*0}$ ($\tan^2\alpha\ll 1$)
and the charmonium. For more general molecules,
it will determine how much asymmetry exists between the 
$D^0\overline{D}^{*0}$ and the $D^+\overline{D}^{*-}$
components inside the molecule $X(3872)$. The BaBar 
Collaboration recently provided\cite{Ricciardi} 
\begin{equation}
     B(B^0\to K^0X(3872))/B(B^+\to K^+X(3871))
        = 0.50 \pm 0.30 \pm 0.05, \label{b} 
\end{equation}  
which is still inconclusive in distinguishing between
the molecule and the charmonium. A smaller statistical 
error will decisively rule out the molecule. 

Although we definitely favor the charmonium over the
molecule, even the charmonium with mixing still has potential
difficulties. In addition to the molecule and the charmonium, 
many other models have been proposed\cite{molecule3}.
However, there is less handle to extend our semiquantitative 
discussion to those models. A common difficulty for them is 
that the production rate is likely much lower than the 
experimental observation since they are generally objects 
of large spatial spread.  The branching fraction ratio of
Eq. (\ref{b}) is useful in distinguishing among them. This 
ratio is unity for the hybrid ($c\overline{c}g$) and the 
glueball ($gg$). It is far away from unity for the 
$cu\overline{cu}$.

\acknowledgements

This work was supported in part by the Director, Office of Science, 
Office of High Energy and Nuclear Physics, Division of High Energy 
Physics, of the U.S. Department of Energy under contract 
DE--AC02--05CH11231 and in part by the National Science Foundation 
under grant PHY-0098840.

\end{document}